# A self-consistent assessment of multi-dimensional fitness of cities

Anand Sahasranaman[1,2,*] and Henrik Jeldtoft Jensen[1,3]

[1]Centre for Complexity Science, Dept of Mathematics, Imperial College London, London SW72AZ, UK.

[2]Division of Mathematics and Computer Science, Krea University, Sri City, AP 517646, India.

[3]Institute of Innovative Research, Tokyo Institute of Technology, 4259, Nagatsuta-cho, Yokohama 226-8502, Japan.

[*] Corresponding Author. Email: anand.sahasranaman@krea.edu.in

**Abstract:**

Given the importance of urban sustainability and resilience to the future of our planet, there is a need to better understand the interconnectedness between the social, economic, environmental, and governance outcomes that underline these frameworks. Here, we propose a synthesis of the independent scientific frameworks of economic complexity and urban scaling into a consistent mechanism – termed '*city complexity*' - to measure the fitness of cities across multiple dimensions. Essentially, we propose the use of urban scaling as the basis to construct and populate a bipartite city-outcome matrix, whose entries are the deviations from scaling law for a given set of urban outcomes. This matrix forms the input into the economic complexity methodology, which iterates over a pair of coupled non-linear maps, computing fitness of cities and complexity of outcomes. We test our algorithm with data from American cities and find that the emergent city fitness measure is consistent with desired behavior across the set of outcomes studied. We also find temporal evolution of city fitness and outcome complexity to be in agreement with theoretical expectation. Overall, these findings suggest that the city complexity mechanism proposed here produces a robust measure of fitness and can be applied for any set of diverse outcomes, irrespective of the specifics of national urban contexts.

## 1. Introduction:

We live in an increasingly urban world, with projections that the share of global urban population is expected to rise to 66% by 2050 [1]. This means that close to 2.5 billion people are expected to migrate to cities between 2014 and 2050, primarily in the developing countries of Asia and Africa [1]. As cities have become central to human existence, there has been increasing focus on notions of urban sustainability and resilience [2-4]. In fact, one of the stated objectives under the United Nations' Sustainable Development Goals (SDGs) is the creation of sustainable cities and communities, which encompasses targets under the multiple dimensions of basic service availability, housing and transport, social development, economic performance, environmental management, and planning processes [2,3]. The increasing impact of climate change has also led to concurrent developments in the notion of urban resilience, with the City Resilience Framework (CRF) defining urban resilience through a basket of 12 indicators under the four broad dimensions of health and well-being, economy and society, infrastructure and environment, and leadership and strategy [4]. In addition to these frameworks for resilience and sustainability, there are multiple other assessment methodologies comprised of indicators across various dimensions of urban experience [5-8]. However, despite a plethora of such assessment methodologies, most extant methods do not consider or explain the interdependencies between the many dimensions considered in each methodology [5].

In this work, we propose a self-consistent, empirically grounded methodology, underpinned by a consideration of the interconnectedness between outcomes across multiple dimensions, to assess the comparative 'fitness' of cities. This measure of 'fitness' could apply to urban sustainability, urban resilience, or any other comparative assessment of cities across a range of outcomes. The methodology we propose - *city complexity* - synthesizes the frameworks of economic complexity [9-11,13,14] and urban scaling [18,19,22-24] to produce both a measure of comparative 'fitness' of cities and an assessment of the relative 'complexity' of outcomes considered.

Economic Complexity was proposed as a framework to assess the productive capability of nations [9,10]. Essentially, it posits that nations have underlying capabilities that are not tradable and difficult to measure such as infrastructure, legal frameworks, and human capital endowments, and that these capabilities determine the kinds of products they are able to produce – that is, the product baskets of nations are reflections of sets of capabilities they possess. Hidalgo and Hausmann proposed Economic Complexity as a method of using a nation's export product basket to infer the exclusivity and diversity of the country's underlying capabilities [9]. They argued that this was possible if we consider the bipartite graph of countries and the products they export (on which data is available) to be part of a larger tripartite graph with an intermediate layer of capabilities (that are difficult to characterize and measure) – meaning that countries link to capabilities they possess and products link to the capabilities required to produce them. Essentially, this translated into countries producing those products for which they have the requisite capabilities. They found empirical validation in the country-product matrix constructed using global trade data (ordered by fitness of nations and complexity of products), which describes a triangular shape, supporting the contention that nations produce all products within a certain complexity limit. Therefore, a product produced by a large number of countries implies a low complexity product, while a product produced only by very high fitness countries implies a high complexity product [9,10]. Using a simple model of the tripartite graph, they were also able to theoretically validate the expectation that countries with greater capabilities were more diversified and produced less ubiquitous products [9]. This work was extended further by Tacchella et. al [11] who proposed a statistical approach based on coupled non-linear maps, inspired by the PageRank algorithm [12], to iteratively arrive at self-consistent measures of fitness of nations. This mechanism was also found to generate good long-term economic growth predictions [13,14].

Given the robust theoretical and empirical foundations of economic complexity, we seek to extend this framework to devise the notion of *city complexity* to assess fitness of cities. For city complexity, we propose that the bipartite graph of cities and their outcomes across multiple dimensions (such as social, economic, environmental etc.) reflects a contraction of the tripartite graph linking cities to their underlying capabilities (such as governance, finance, institutions, infrastructure etc.) and outcomes to capabilities required to achieve them. Figure 1 presents a visual representation of the equivalence between the empirically observed bipartite graph linking cities to outcomes and the underlying tripartite graph model.

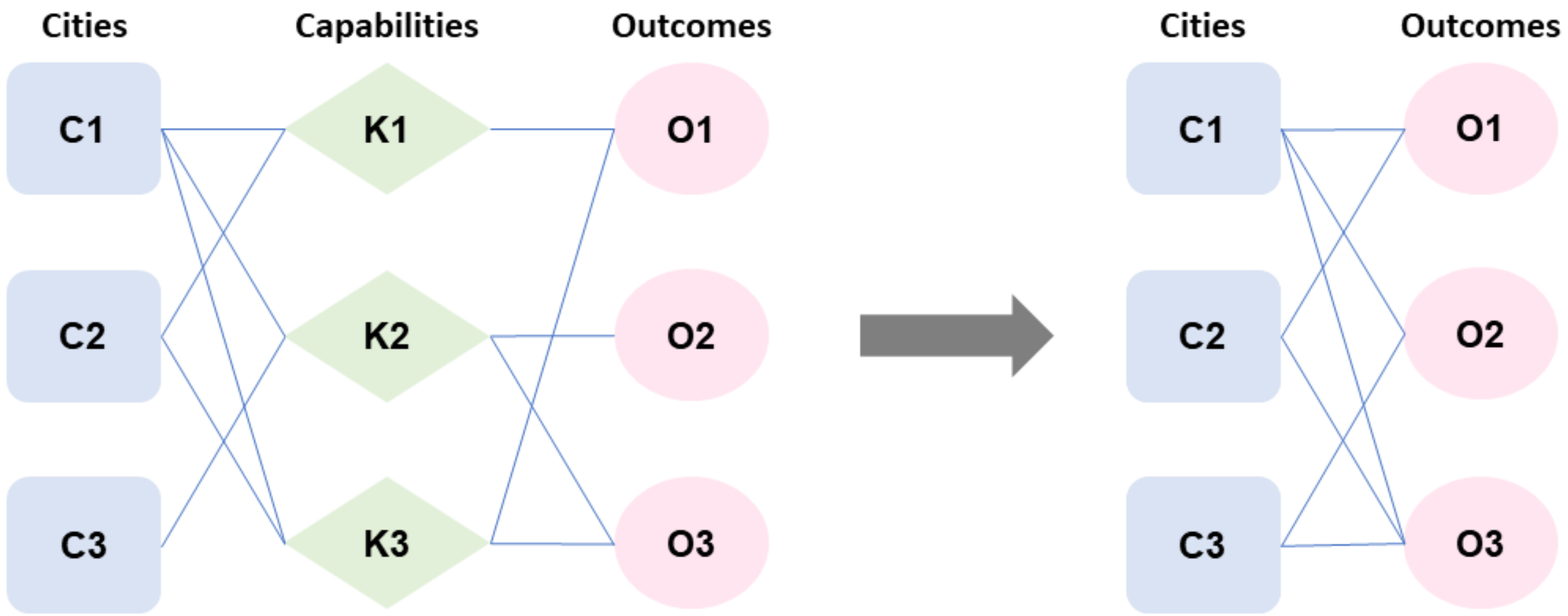

***Figure 1:*** **Decomposition of model tripartite graph to actual bipartite graph:** The tripartite graph on the left-hand side represents the underlying model linking cities to capabilities possessed and outcomes to capabilities required. The equivalent bipartite decomposition on the right-hand side represents the actual data linking cities and outcomes.

In this context, the complexity of an outcome is a measure of the difficulty inherent in cities being able to generate the combination of underlying capabilities required to produce that outcome, and the fitness of a city is a measure of its performance across outcomes of different complexities. Given this construct, we propose that the 'fitness' of cities ($F$) and the 'complexity' of outcomes ($Q$) are obtained as the unique fixed point of the iteration of non-linear coupled maps, as conceived by Tacchella et. al [11]. Specifically, this means that at any given iteration $n$, we estimate two sets of variables, namely the fitness of each city $c$ ($F_c{}^n$) and the complexity of each outcome $o$ ($Q_o{}^n$), as follows:

$$F_c^n = \frac{\sum_o M_{co} Q_o^{n-1}}{<\sum_o M_{co} Q_o^{n-1}>_c} \tag{1}$$

$$Q_o^n = \frac{\frac{1}{\sum_c \frac{M_{co}}{F_c^{n-1}}}}{<\frac{1}{\sum_c \frac{M_{co}}{F_c^{n-1}}}>_o}, \tag{2}$$

where $M_{co}$ is the entry for city $c$ and outcome $o$ in the bipartite city-outcomes matrix ($M$). The initial conditions for the iterations are:

$$F_c^0 = 1 \forall c;\ Q_o^0 = 1 \forall o,$$

The Fitness of a city ($F_c$) is proportional to the linear sum of the complexity of its outcomes (Eq. 1), while Complexity of an outcome ($Q_o$) weights the Fitness of cities that produce the outcome in a non-linear way (Eq. 2), so that the complexity of an outcome is bound by the Fitness of less competitive cities that manifest them. At the end of the iterative process, we obtain a rank ordering of cities by fitness and of outcomes by complexity. It has also been numerically shown that the fixed point of Eqs. 1 and 2 exists and is unique, meaning that the result is independent of the choice of initial conditions [15]. The convergence of paths with different initial conditions is found to be exponential, with dependence on matrix size [15]. The city complexity methodology ensures that the weightage given to any particular outcome in the computation of fitness is solely dependent on the difficulty in achievement of that outcome (which itself is a function of the fitness of cities that manifest the outcome), and this distinguishes our mechanism from other multi-dimensional indices and methods that have a priori assignment of weights for different outcomes (including for the UN's Sustainable Development Goals (SDG) Index [16,17]).

The central question that remains unresolved is the mechanism to construct the bipartite matrix $M$. In the case of economic complexity, the widely used measure (in economics) of Revealed Comparative Advantage (RCA) was used to populate the matrix [18]. In the case of city complexity, we propose to use urban scaling as the basis to construct the matrix as it provides us a scientifically robust framework to assess the performance of cities across a range of attributes. Urban scaling [19,20,23-25] uses population size as the basis for isolating general agglomeration effects, with any urban indicator $Y_i(t, N_i)$ for city $i$ at time $t$ described as:

$$Y_i(t, N_i) = Y_0(t) N_i(t)^{\beta} e^{\xi_i(t)} , \tag{4}$$

where, $N_i(t)$ is the population of the city at time $t$, $Y_0(t)$ is the systematic change of the indicator under consideration, and $\xi_i(t)$ represents the non-systematic or idiosyncratic city-specific deviation from the scaling law. $\beta$ is the elasticity of the urban indicator relative to population at $t$ and urban scaling theory predicts that $\beta$ falls into three universality classes for different types of urban indicators: $\beta \simeq 7/6$ indicating superlinear scaling for socioeconomic parameters, $\beta \simeq 5/6$ indicating sublinear scaling for public infrastructures, and $\beta \simeq 1$ indicating linear scaling for infrastructures representing individual needs [19]. Urban scaling theory predicts these average spatial, social, and infrastructural properties of cities as scaling relationships based on a few basic principles operating locally: mixing populations, incremental network growth, bounded human effort, and proportionality of socioeconomic outputs to social interactions [20]. These urban agglomeration properties are sometimes not evinced when the city boundaries considered in the analysis do not correspond to the functional definition of the city (i.e. an Urban Agglomeration comprising together both places of residence and work) [21,22]. The robustness of urban scaling theory has been confirmed by empirical validation of scaling behavior in Urban Agglomerations across cities in Europe, USA, China, Brazil, Mexico, and India [19,23-25].

The scaling law - $\ln Y_0(t) + \beta \ln N_i(t)$ – essentially represents the mean-field behavior of cities for a given outcome at time $t$, and we propose to use a measure of deviation from scaling law as the basis for populating the city-outcomes matrix $M$. This deviation is in essence a measure of the city-specific component of performance, beyond the systemic change for all cities represented by the scaling law. Depending on the outcome under consideration, desired behavior (representing higher fitness) could be either overperformance of the scaling law (as for example in the case of Gross Domestic Product or number of patents) or underperformance of the scaling law (for outcomes such as crime, poverty, or unemployment). In case desired behavior on a particular outcome $o$ is defined as overperformance of scaling law, we define the entry $M_{co}$ (in matrix M) for any city $c$ for outcome $o$ as:

$$M_{co} = \frac{\ln Y_c(t,N_c)}{\ln Y_0(t) + \beta \ln N_c(t)} \tag{5a}$$

Alternatively, if desired behavior is represented by underperformance of the scaling law, then:

$$M_{co} = \frac{\ln Y_0(t) + \beta \ln N_c(t)}{\ln Y_c(t,N_c)} \tag{5b}$$

Once *M* is populated using Eqs. 5a and 5b, we can execute the iterative procedure outlined earlier (Eqs. 1 and 2) to compute city complexity. Overall, this synthesis of the independent frameworks of economic complexity and urban scaling enables us to construct a systematic, self-consistent method to assess multi-dimensional fitness of cities. We now seek to validate the model using data from American cities.

## 2. Results and Discussion:

In general, indices of urban sustainability encompass four broad dimensions – economic, social, environmental, and governance [2,3]. We find data on 11 outcome parameters across these four dimensions for Urban Agglomerations in the USA: Gross Domestic Product (GDP), USPTO utility patent count, poverty headcount, unemployed headcount, population with educational attainment of bachelors' and higher, violent crime count, number of housing units without complete kitchen facilities, number of housing units without complete plumbing facilities, person-days of good Air Quality Index (AQI), population taking public transport, and total time to work. It is of course possible to find a more comprehensive or varied set of outcomes to assess sustainability, but our intention here is to simply test our methodology on a set of defined outcomes. For a more detailed description of data and methods, refer Appendix A.

Given this specific set of outcomes, we define over performance of scaling law as desired behavior for outcomes of GDP, patent count, educational attainment, person-days of good AQI, and population taking public transport; while underperformance from scaling law is desired behavior for poverty, unemployment, violent crime, lack of complete kitchen facilities, lack of complete plumbing facilities, and time taken to work. We begin with an assessment of 262 MSAs that report data across all the 11 chosen outcomes for the year 2015. Table 1 presents the scaling exponents for all the outcomes, which are broadly in line with expectations from urban scaling theory. A detailed discussion on these scaling relationships in presented in Appendix B.

| **Rank (ordered by $Q_o$)** | **Outcome** | **Scaling exponent (β)** | **95% Confidence Interval** | ***R*-squared** |
|---|---|---|---|---|
| 1 | USPTO utility Patents | 1.35 | 1.23 – 1.47 | 0.67 |
| 2 | Population taking public transport | 1.42 | 1.34 – 1.51 | 0.80 |
| 3 | Educational attainment count (bachelor's and higher) | 1.10 | 1.07 – 1.13 | 0.95 |
| 4 | GDP | 1.10 | 1.07 – 1.13 | 0.96 |
| 5 | Person-days of good AQI | 0.81 | 0.77 – 0.85 | 0.86 |
| 6 | Poverty headcount | 0.96 | 0.93 – 0.99 | 0.94 |
| 7 | Total time to work | 1.10 | 1.08 – 1.11 | 0.99 |
| 8 | Unemployment headcount | 1.03 | 1.01 – 1.06 | 0.95 |
| 9 | Housing units with lack of complete kitchen facilities | 0.95 | 0.90 – 0.99 | 0.89 |
| 10 | Violent crime count | 1.13 | 1.08 – 1.18 | 0.87 |
| 11 | Housing units with lack of complete plumbing facilities | 0.96 | 0.91 – 1.02 | 0.82 |

***Table 1:*** **Scaling exponents for all outcomes and rank ordering of outcomes by revealed complexity:** The scaling exponents obtained are in broad agreement with theoretical expectation. The rank ordering of outcomes by $Q_o$ is a measure of the comparative difficulty inherent in achieving an outcome relative to other outcomes.

Given the scaling laws for different outcomes, the bipartite city-outcome matrix (*M*) is populated in accordance with Eqs. 5a and 5b. At the end of the iteration of the two non-linear coupled equations (Eqs. 1 and 2), an assessment of the comparative complexity of the 11 outcomes indicates that the most 'complex' outcomes are utility patents (tracking technological innovation) and public transport use, while the least 'complex' are access to housing with complete plumbing facilities and violent crime. The notion of complexity here is best understood in terms of combinations of underlying city capabilities required to achieve a given outcome. The higher ranked outcomes are essentially those derived by combinations of harder to develop capabilities, while the lower ranked ones are achieved using more ubiquitously available capabilities across cities. Given this understanding and also considering the unlikelihood of changes in the set of underlying capabilities contributing to a given outcome over the timeframe of a few years, we would expect that a time-series of outcome rankings by complexity would reveal a fairly unchanging rank order over short time scales (years), and that only over longer time scales (decades) would we expect to see any significant ranking change. The coupled nature of the relationship between complexity and fitness also implies that cities starting at the highest levels of fitness would, with high probability, retain their position amongst the most fit cities over short time scales due to the availability of harder to develop capabilities, but as the capability set declines going down the ladder of fitness we would expect increasing probability of movement in the rankings of cities.

We now test the performance of the Fitness measure ($F_c$) against our definition of the desired behavior on each of the 11 parameters. In order for our measure of fitness to be robust, we would expect that it tracks desired behavior across most, if not all, outcomes. This would mean that $F_c$ ought to describe an increasing relationship with outcomes of GDP, patent count, educational attainment, person-days of good AQI, and population taking public transport, and a decreasing relationship with poverty, unemployment, violent crime, lack of complete kitchen facilities, lack of complete plumbing facilities, and time taken to work. Figure 1 plots these relationships and reveals that $F_c$ tracks desired behavior in all cases, except for time taken to work which shows a slightly positive relationship against an expected negative relationship. An analysis of the relationship between fitness and outcomes for the period 2011 - 2015 reveals that in each of these years, fitness tracks desired behavior across outcomes (Appendix C presents these results). Overall, the emergent relationships between fitness and outcomes are in keeping with our expectations, suggesting that the Fitness measure ($F_c$) is robust and also indicating that the methodology would be robust to a diverse choice of urban outcomes. Empirical tests with urban data from other nations and longer time series would be essential for further validation of the measure. Additionally, specific times of socioeconomic stress which have a significant impact on urban life (such as economic crises, social unrest etc.) could result in greater churn in both short-term city fitness and outcome complexity measures due to the possibility of rapid, as well as differential, change in outcomes across cities.

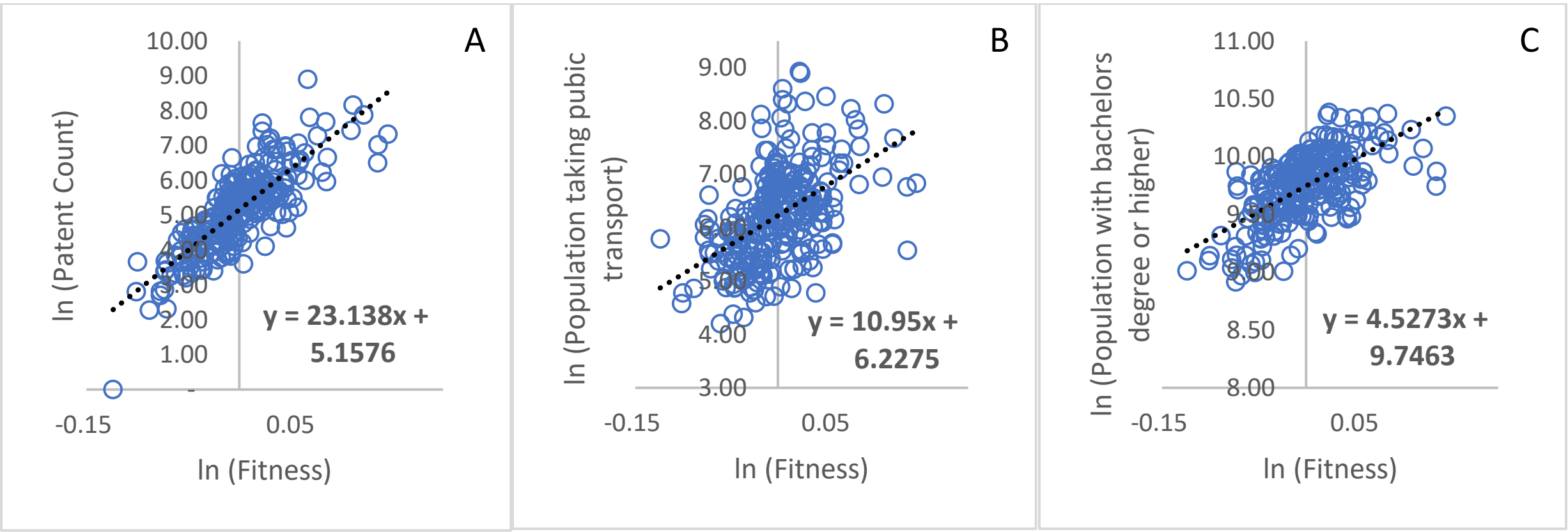

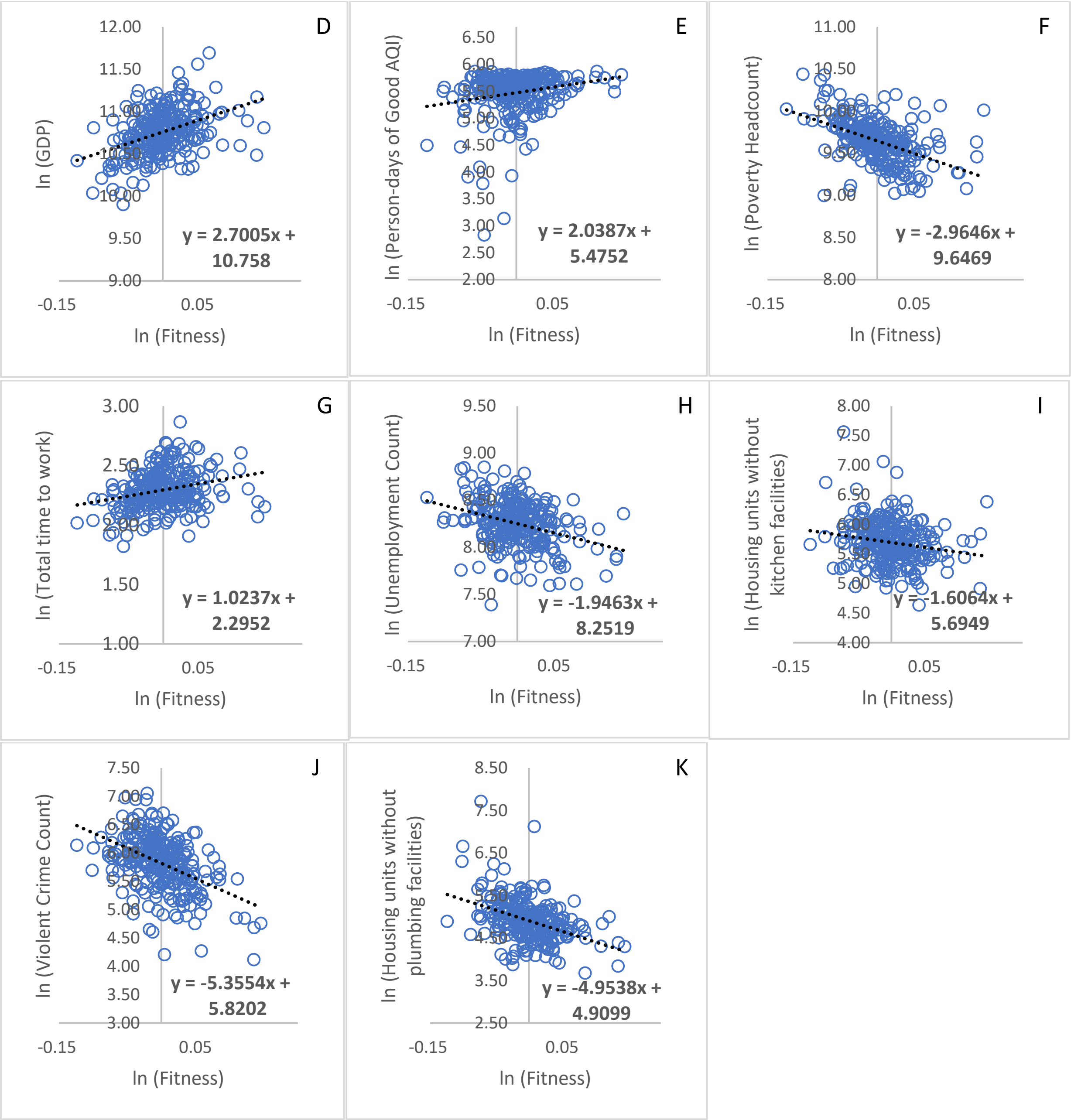

***Figure 2*: Fitness relationship with outcomes:** Fitness describes increasing relationship with the following outcomes, as expected: A: ln (Fitness) v ln (Patent Count). B: ln (Fitness) v ln (Population taking public transport). C: ln (Fitness) v ln (Population with bachelor's degree or higher). D: ln (Fitness) v ln (GDP). E: ln (Fitness) v ln (Person-days of Good AQI). Fitness describes decreasing relationship with the following outcomes, as expected: F: ln (Fitness) v ln (Poverty Headcount). H: ln (Fitness) v ln (Unemployment Count). I: ln (Fitness) v ln (Housing units without kitchen facilities). J: ln (Fitness) v ln (Violent Crime Count). K: ln (Fitness) v ln (Housing units without plumbing facilities). Fitness describes relationship contrary to expectation with the following outcome: G: ln (Fitness) v ln (Total time to work) describes a positive relationship as against an expected negative relationship. In summary, it emerges that Fitness describes relationships with urban outcomes that are in line with expectations in 10 out of 11 cases.

Next, we seek to understand the temporal evolution of city fitness and outcome complexity measures. We analyze the time-series for the period 2011-2015 (which gives us a 5-year dataset of 178 cities for analysis, Appendix A), and execute the city complexity algorithm - computing the fitness of cities and complexity of outcomes for each year by constructing the city-outcome matrix based on the scaling relationships described by the data for that year. Figure 3a displays the outcomes on temporal evolution of city fitness. We find that, on average, a city's fitness ranking changes by 15.9 (or 8.9%) over the 5-year period. While this is a mean-field description of change, we also find a systematic relationship between the magnitude of rankings change and the initial fitness of cities, with high initial fitness corresponding to lower magnitude of average ranking change over time. To quantify this behavior more systematically, we stratify cities into quartiles based on fitness rankings (45 cities in each of the first 3 quartiles and 43 in the 4th quartile), and find that the revealed probability of city rankings changing by 10 places or more over the 5-year period is 0.33 for the first quartile, 0.58 for the second quartile, 0.51 for the third quartile, and 0.35 for the fourth quartile (Figure 3b). This emergent outcome is in keeping with the expectation that the harder to develop capability set of the highest fitness cities would enable them to retain their fitness rankings with higher probability over time, but it also reveals that the least fit cities display similar levels of stickiness to their rankings – indicating the possibility of 'low fitness' traps, which reflect the difficulty that very low fitness cities face in developing more complex capabilities. Intermediate quartiles, as expected, show much higher probability of rankings change.

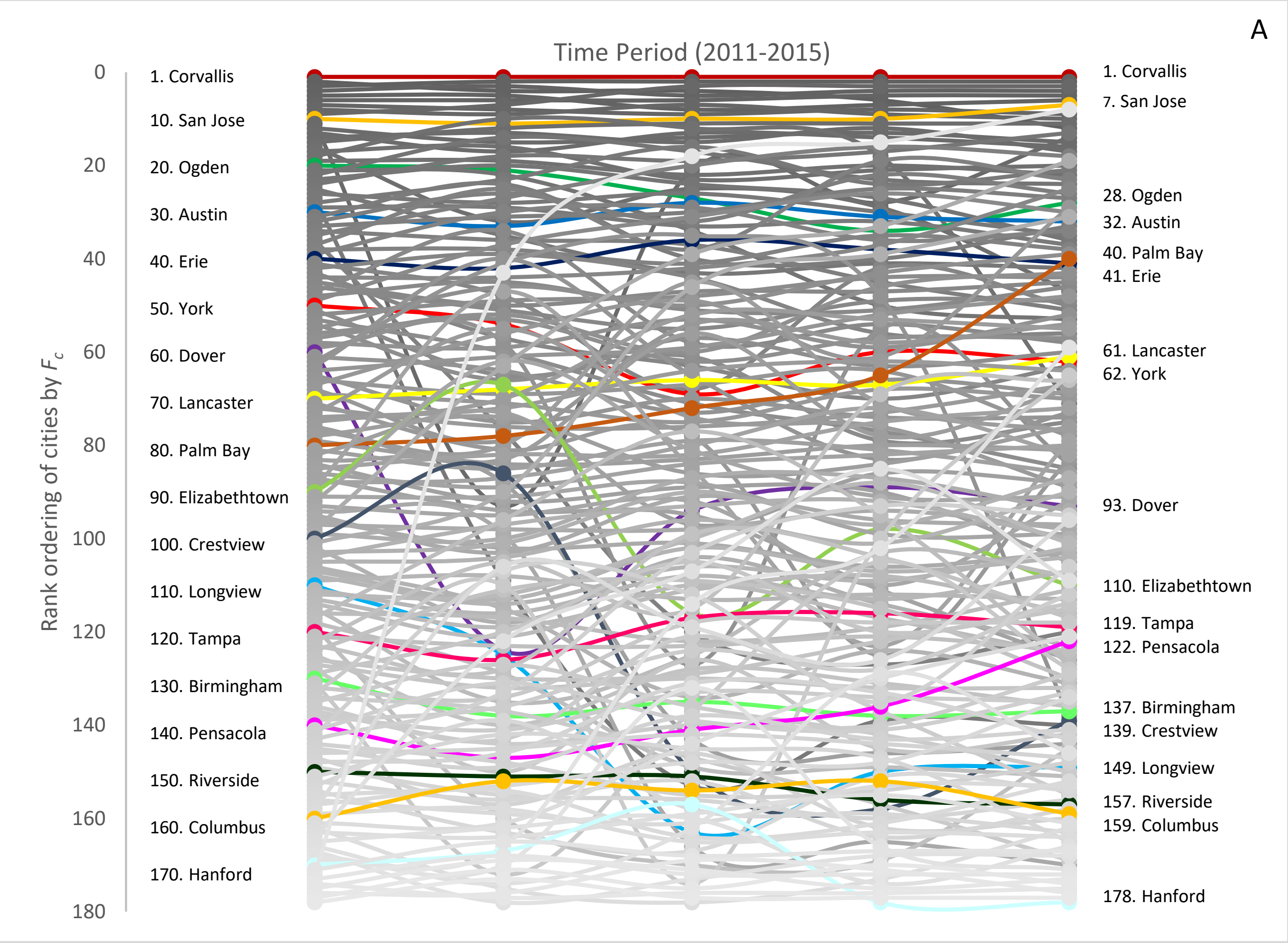

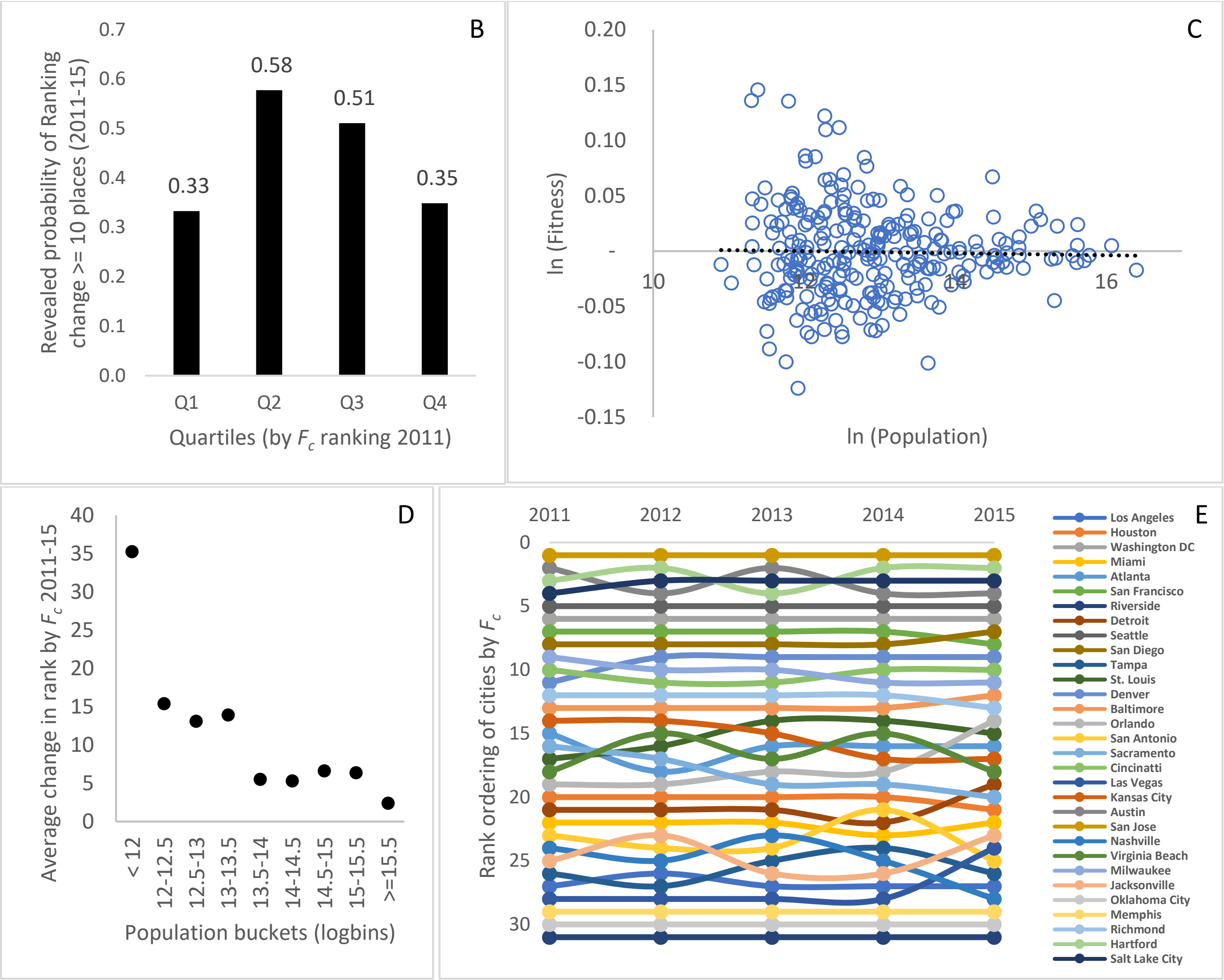

***Figure 3*: Temporal evolution of city fitness ($F_c$):** A: Rank ordering of cities by $F_c$ between 2011-2015: The temporal evolution of fitness rankings of cities suggests historical path-dependence. B: Revealed probability of ranking change >= 10 positions between 2011-15 v Quartiles (ordered by $F_c$ ranking 2011): Cities are grouped into quartiles by initial $F_c$ rankings (in 2011) and the revealed probability of rankings change by 10 spots or greater over the 5-year period shows that the highest fitness quartile and lowest fitness quartile show significantly greater stickiness to rankings that the intermediate quartiles. C: ln (Population) v ln (Fitness): Fitness shows no discernible relationship with population. D: Average change in $F_c$ rank between 2011 and 2015 v. Population (grouped in equi-sized logarithmic bins): Larger cities, on average, show significantly lesser change in fitness rankings than smaller cities. E: Rank ordering of million-plus cities by $F_c$ between 2011-2015: This plot further illustrates temporal path-dependence in the evolution of $F_c$.

Assessing the rankings, we find that the top cities by $F_c$ rankings are all small and medium sized cities such as Corvallis, Or., Bremerton, Wa., and Iowa City, Ia. (and they retain their high ranking over the 5-year period) and the top ranked million-plus city over this time period is San Jose, Ca. This suggests that the evolution of $F_c$ exhibits temporal path dependence and also the possibility of population being a predictor of fitness. However, we find no systematic relationship between population and $F_c$, clearly indicating that the size of cities is no indicator of their multi-dimensional fitness (Figure 3c). Population does, however, appear to be a predictor of the magnitude of fitness ranking changes over time, with larger cities, on average, showing lower ranking change than smaller cities (Figure 3d). This is indicative of the difficulty in inducing change in outcomes of large systems when compared to smaller systems. Essentially, this finding suggests that larger cities (pop. > 725,000) displaying poor overall performance on fitness will find it difficult to significantly change their ranking positions in short time periods (5 years as in this case). In the same time frame, smaller cities (pop. < 725,000), on average, can show significant change in fitness ranking. In order to further explore the dynamics of fitness change in large cities, we do a comparative analysis of the 31 largest cities in our dataset with a population of over 1 million in 2015 (Figure 3e). Figure 3e evinces the clear temporal path dependence of Fitness ($F_c$), with an average change in 5-year ranking of 1.4 (or 4.8%), suggesting that significant improvement in the fitness of cities would require longer time horizons - to develop requisite capabilities that consequently enable achievement of more complex outcomes. This is in agreement with the finding that larger cities exhibit greater stickiness in rankings over time (Figure 3c). Amongst the large cities, San Jose, Ca., Austin, Tx., Hartford, Ct., Salt Lake City, Ut., and Seattle, Wa., have the highest fitness scores (and they keep their position in the top 5 ranks through the 5-year period), while cities like Miami, Fl., Tampa, Fl., Los Angeles, Ca., and Riverside, Ca. find themselves in the bottom third of fitness rankings through the timeframe of analysis. The complete Fitness ranking of American cities is presented in Appendix D.

When we look at the evolution of complexity of outcomes over time, we find that there is almost no change in the relative rankings of outcome complexity. As Figure 4 reveals, there is no change in ranking of the top 8 attributes between 2011 and 2015, and the only change is a one position swap in rankings between two of the lesser complex outcomes (ranked 9 and 10). This result confirms our expectations at the outset that complexity ranking of outcomes would not exhibit change over short time scales because of the low probability of any change in the underlying capabilities required to achieve a given outcome and the progressive difficulty faced by cities in developing harder to achieve capabilities.

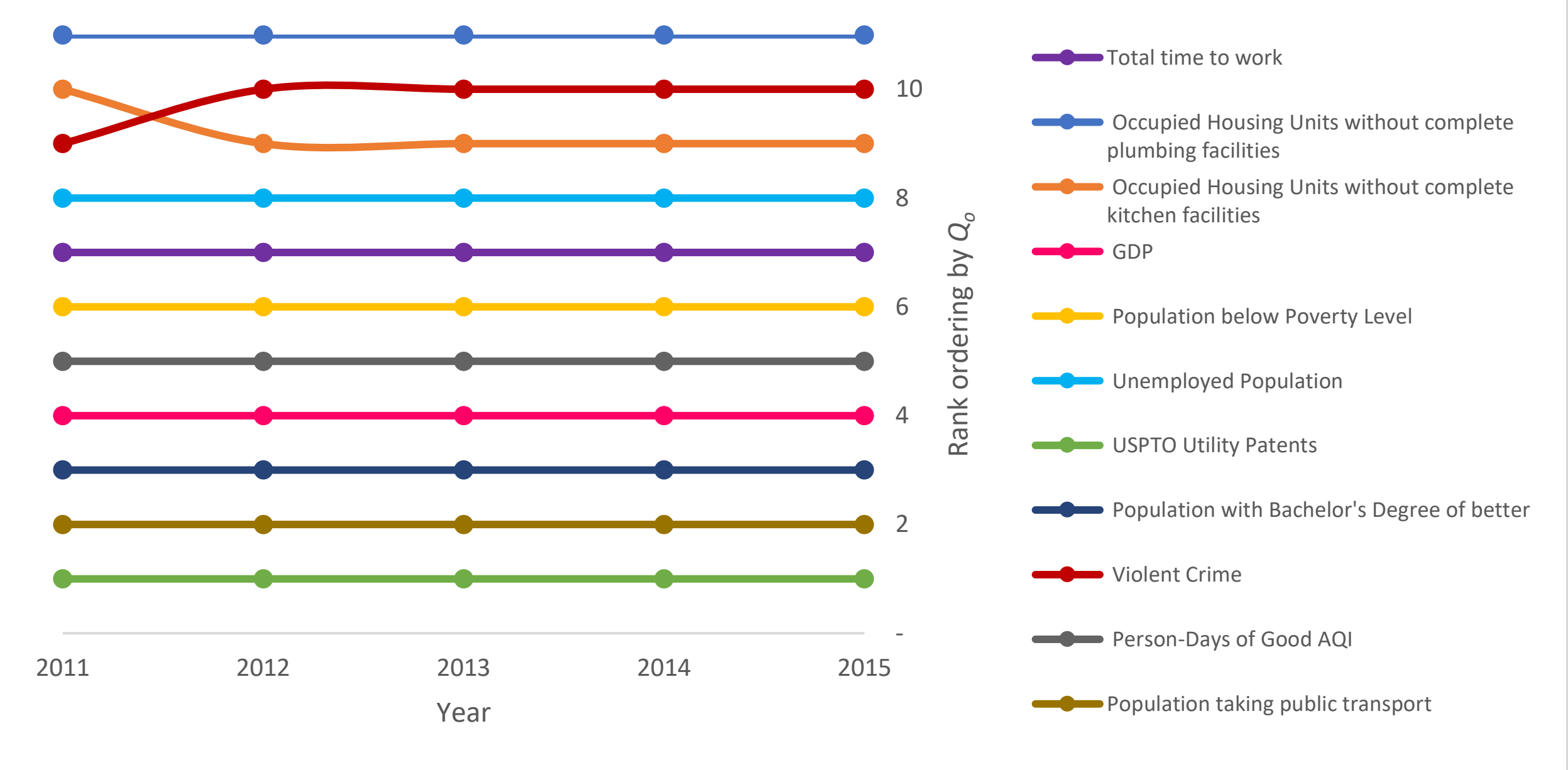

***Figure 4*: Rank ordering of outcomes by $Q_o$ between 2011-2015:** Temporal evolution of outcome complexity shows almost no change in relative rankings between 2011 and 2015.

This analysis of the robustness of the fitness measure as well as the study of temporal evolution of fitness and complexity suggests that our proposed mechanism of city complexity offers an empirically grounded and theoretically robust algorithm to assess multi-dimensional fitness of cities.

## 3. Conclusion:

In this work, we attempt to create a theoretically sound and empirically grounded mechanism called *city complexity* to assess the fitness of cities. We synthesize the independent scientific frameworks of economic complexity and urban scaling into a consistent method to measure the fitness of cities based on their performance on a range of multi-dimensional outcomes. Essentially, we propose the use of urban scaling as the basis to populate a city-outcome matrix, which forms the input into the economic complexity methodology of iterating over a pair of coupled non-linear maps so as to compute the fitness of cities and complexity of outcomes. We test this city complexity algorithm with data from American cities and find that the emergent city fitness measure is consistent with desired behavior across the set of outcomes studied. We also find that while population is not a predictor of fitness, it is indeed a predictor of the extent of change in fitness (measured by ranking change) of cities. The magnitude of ranking change in fitness of cities is found to decrease with increasing population. We study the temporal evolution of city fitness and outcome complexity for the period 2011-2015, and find that the relative rankings of complexity of outcomes remains unchanged over time, as per expectations. We also find that cities starting at high levels of fitness tend to display lesser change in fitness rankings than those starting at lower fitness. This finding is again in agreement with theoretical expectation. We also find that the lowest fitness cities show similar stickiness to their rankings, suggesting the possibility of 'low fitness' traps. These findings suggest that the city complexity mechanism proposed here produces a robust measure of fitness and complexity.

There are a wide range of indicators and methods to assess city sustainability and resilience being used by governments, NGOs, research institutions, and private organizations around the world. This proposal for city complexity, we believe, offers a completely self-consistent, data-driven, and theoretically grounded methodology to evaluate city fitness - a mechanism that is robust to varying specifications of outcomes and implementable irrespective of the particularities of urban or national context.

# Supplementary Information

**A self-consistent mechanism to assess multi-dimensional fitness of cities**

**Anand Sahasranaman, Henrik Jeldtoft Jensen**

## Appendix A: Data and Methods

We obtained data at the level of Metropolitan Statistical Authority (MSA) from the following sources:

*The American Community Survey* (https://factfinder.census.gov/faces/nav/jsf/pages/index.xhtml) provides data on a range of attributes including social, economic, demographic, and housing characteristics at many different levels of geographic granularity. Specifically, for our analysis, the following data was obtained was the level of the MSA (for the years 2011 to 2015):

1. Total population
2. Occupied housing units lacking complete plumbing facilities
3. Occupied housing units lacking complete kitchen facilities
4. Workforce size
5. Unemployment rate
6. Population fraction with income below poverty level
7. Population over 25
8. Fraction of population over 25 with bachelors' degree or better
9. Population taking public transport
10. Population commuting to work
11. Mean time to work

Apart from the following 3 measures: housing units lacking complete kitchen facilities, housing units lacking complete plumbing facilities, and population taking public transport, which we used directly for our analysis, we computed other input data as follows:

a. Unemployed headcount = Workforce size * Unemployment rate
b. Poverty headcount = Total population * Population fraction with income below poverty level
c. Population with education of bachelors' or higher = Population over 25 * Fraction of population over 25 with bachelors' degree or better
d. Total time to work = Population commuting to work * Mean time to work

The *Federal Bureau of Investigation* produces an annual publication titled *Crime in the United States* (https://www.fbi.gov/services/cjis/ucr/publications) that details a range of crime statistics. Specifically, for our analysis, we look at the statistics on violent crime, defined as "..composed of four offenses: murder and nonnegligent manslaughter, rape, robbery, and aggravated assault." We use violent crime data for the years 2011-2015 at the level of MSAs. Violent crime is reported as - violent crime rate (per 100,000 inhabitants). Given this data, the input into our analysis is computed as:

- Violent Crime Count = Violent crime rate * 100,000

Data on patent statistics is available from *United States Patents and Trademarks Office* (USPTO - https://www.uspto.gov/web/offices/ac/ido/oeip/taf/reports_cbsa.htm). A complete listing of utility patent counts by MSA is available for 2011-2015, and this is a direct input into our analysis.

Data on Gross Domestic Product (GDP) is available from the *Bureau of Economic Analysis* (BEA - https://apps.bea.gov/regional/histdata/). This data is available disaggregated at the level of MSAs for the period 2011-2015 and is a direct input into our analysis.

Data on the Air Quality Index (AQI) is available from the *Environmental Protection Agency* (EPA- https://aqs.epa.gov/aqsweb/airdata/download_files.html#Annual). Specifically, for our analysis, we get the following data:

1. Number of days with Good AQI
2. Days with AQI measurement

With these two data points available at the level of MSAs, we consider only those cities in our analysis that have days with AQI measurement >= 300. If this condition is satisfied, we compute the following outcome, which is an input into our analysis:

- Person-Days of Good AQI = Total Population * 365 * Number of days with Good AQI / Days with AQI measurement

The dataset for analysis requires that all MSAs considered have data across all 11 outcomes. This has meant that some large metropolises, such as New York City, Boston, and Chicago, do not feature in the final dataset on account of absence of data on certain outcomes. Overall, the 2015 dataset contains 262 cities, each with a complete set of 11 measured outcomes. The time series (2011-2015) dataset comprises 178 cities.

**Appendix B: Scaling Analysis**

An analysis of the scaling exponents (Table B1) reveals that the socioeconomic outcomes of GDP, utility patent count, educational attainment count, violent count, total time to work, population taking public transport all scale superlinearly with city size. This is in agreement with empirical observation from around the world as well as urban scaling theory. Poverty count is found to scale sublinearly with population, which is arguably consistent with the corollary expectation of superlinear scaling of economic income. The environmental attribute of Person-days of good AQI scales sublinearly with population, which is perhaps a reflection of the fact that large cities have become centers of the new economy based on technology and other services, while manufacturing industries are more prevalent in smaller urban centres. Finally, the household level personal infrastructures of complete kitchen and plumbing facilities would be expected to scale linearly (if universal access to basic services is assumed), but are found to scale sub-linearly, which is an indication that as American cities grow bigger, larger fractions of households do not have accesses to these basic services.

| **Outcomes** | **2015** | **2014** | **2013** | **2012** | **2011** |
|---|---|---|---|---|---|
| USPTO utility Patents | 1.35<br>[1.23 – 1.47] | 1.43<br>[1.28 – 1.57] | 1.45<br>[1.30 – 1.60] | 1.50<br>[1.34 – 1.66] | 1.42<br>[1.27 – 1.58] |
| Population taking public transport | 1.42<br>[1.34 – 1.51] | 1.43<br>[1.31 – 1.54] | 1.42<br>[1.31 – 1.53] | 1.42<br>[1.31 – 1.54] | 1.43<br>[1.32 – 1.54] |
| Educational attainment count (bachelor's and higher) | 1.10<br>[1.07 – 1.13] | 1.11<br>[1.07 – 1.15] | 1.11<br>[1.07 – 1.15] | 1.11<br>[1.07 – 1.16] | 1.11<br>[1.07 – 1.16] |
| GDP | 1.10<br>[1.07 – 1.13] | 1.10<br>[1.06 – 1.13] | 1.10<br>[1.06 – 1.13] | 1.09<br>[1.06 – 1.13] | 1.09<br>[1.05 – 1.12] |
| Person-days of good AQI | 0.81<br>[0.77 – 0.85] | 0.80<br>[0.75 – 0.86] | 0.81<br>[0.75 – 0.86] | 0.79<br>[0.74 – 0.84] | 0.78<br>[0.72 – 0.83] |
| Poverty headcount | 0.96<br>[0.93 – 0.99] | 0.96<br>[0.92 – 1.00] | 0.96<br>[0.92 – 1.00] | 0.96<br>[0.92 – 1.00] | 0.95<br>[0.91 – 0.99] |
| Total time to work | 1.10<br>[1.08 – 1.11] | 1.09<br>[1.07 – 1.11] | 1.09<br>[1.07 – 1.11] | 1.09<br>[1.07 – 1.12] | 1.09<br>[1.07 – 1.12] |
| Unemployment headcount | 1.03<br>[1.01 – 1.06] | 1.05<br>[1.01 – 1.08] | 1.06<br>[1.02 – 1.09] | 1.06<br>[1.02 – 1.10] | 1.06<br>[1.03 – 1.10] |
| Housing units with lack of complete kitchen facilities | 0.95<br>[0.90 – 0.99] | 0.95<br>[0.90 – 0.99] | 0.94<br>[0.89 – 0.98] | 0.94<br>[0.89 – 0.99] | 0.93<br>[0.88 – 0.99] |
| Violent crime count | 1.13<br>[1.08 – 1.18] | 1.13<br>[1.07 – 1.19] | 1.12<br>[1.07 – 1.18] | 1.13<br>[1.07 – 1.19] | 1.13<br>[1.07 – 1.19] |
| Housing units with lack of complete plumbing facilities | 0.96<br>[0.91 – 1.02] | 0.96<br>[0.91 – 1.02] | 0.96<br>[0.88 – 1.03] | 0.96<br>[0.88 – 1.04] | 0.95<br>[0.87 – 1.03] |

***Table B1:* Scaling exponents and 95% Confidence Intervals for all outcomes over the period 2011-2015:** The scaling exponents obtained over time are in close proximity, as we would expect. These exponents are also in broad agreement with theoretical expectation.

Table B1 also reveals that the exponents remain proximate over time, and this is to be expected because cities are unlikely to show dramatic changes in performance across outcomes over spans of a year – therefore ensuring that the outcome distribution remains largely unchanged year on year, yielding similar scaling exponents.

## Appendix C: Robustness of fitness measure

As indicated in the main text, the robustness of the Fitness measure ($F_c$) is contingent on its ability to track desired behavior across all outcomes. Table C1 presents the relationship of with outcomes for the period 2011-2015.

| **Outcomes** | **2015** | **2014** | **2013** | **2012** | **2011** |
|---|---|---|---|---|---|
| USPTO utility Patents | 23.14 | 22.09 | 22.30 | 21.12 | 21.32 |
| Population taking public transport | 10.95 | 10.80 | 10.70 | 8.66 | 9.35 |
| Educational attainment count (bachelor's and higher) | 4.53 | 4.60 | 4.51 | 4.44 | 4.39 |
| GDP | 2.70 | 2.29 | 2.43 | 2.43 | 2.41 |
| Person-days of good AQI | 2.04 | 2.44 | 1.47 | 1.73 | 2.44 |
| Poverty headcount | -2.96 | -3.72 | -3.12 | -2.71 | -3.52 |
| Total time to work | 1.02 | 0.78 | 1.11 | 1.20 | 1.03 |
| Unemployment headcount | -1.95 | -2.21 | -1.42 | -1.25 | -1.75 |
| Housing units with lack of complete kitchen facilities | -1.61 | -1.74 | -1.39 | -1.97 | -2.39 |
| Violent crime count | -5.36 | -5.46 | -4.76 | -4.93 | -4.90 |
| Housing units with lack of complete plumbing facilities | -4.95 | -5.09 | -4.68 | -4.54 | -4.98 |

***Table C1:* Slope of trendline for Fitness v. outcomes 2011-2015:** The background cell color indicates the sign of expected slope (blue = positive, red = negative), while the text color indicates the sign of the realized slope (blue = positive, red = negative). As is apparent, across the timeline from 2011 to 2015, the realized slope for all outcomes is in keeping with expectation, except for Total time to work, which produces a positive slope as against a negative expectation.

Overall, Fitness ($F_c$) is found to track desired behavior for 10 out of 11 outcomes for each of the years under consideration, which suggests that in its current design, $F_c$ is a robust measure of multi-dimensional fitness.

## Appendix D: Fitness results for American MSAs

Table D1 presents the relative fitness rankings of all 178 MSAs considered in the analysis for the period 2010-2015.

| Metropolitan Statistical Area (MSA) | 2011 | 2012 | 2013 | 2014 | 2015 |
|---|---|---|---|---|---|
| Corvallis, OR Metro Area | 1 | 1 | 1 | 1 | 1 |
| Idaho Falls, ID Metro Area | 2 | 4 | 4 | 6 | 12 |
| Bremerton-Silverdale, WA Metro Area | 3 | 2 | 2 | 2 | 2 |
| Iowa City, IA Metro Area | 4 | 3 | 5 | 4 | 4 |
| Charlottesville, VA Metro Area | 5 | 5 | 8 | 7 | 6 |
| Trenton, NJ Metro Area | 6 | 6 | 6 | 5 | 5 |
| Cedar Rapids, IA Metro Area | 7 | 8 | 9 | 8 | 9 |
| Appleton, WI Metro Area | 8 | 9 | 7 | 9 | 10 |
| Bloomington, IN Metro Area | 9 | 7 | 3 | 3 | 3 |
| San Jose-Sunnyvale-Santa Clara, CA Metro Area | 10 | 11 | 10 | 10 | 7 |
| Williamsport, PA Metro Area | 11 | 93 | 21 | 28 | 14 |
| Madison, WI Metro Area | 12 | 15 | 14 | 13 | 15 |
| Bridgeport-Stamford-Norwalk, CT Metro Area | 13 | 17 | 13 | 14 | 16 |
| Manchester-Nashua, NH Metro Area | 14 | 13 | 15 | 12 | 17 |
| Fort Collins, CO Metro Area | 15 | 12 | 11 | 11 | 11 |
| Sioux Falls, SD Metro Area | 16 | 32 | 25 | 29 | 47 |
| Napa, CA Metro Area | 17 | 16 | 23 | 21 | 20 |
| Owensboro, KY Metro Area | 18 | 10 | 12 | 18 | 71 |
| Kennewick-Richland, WA Metro Area | 19 | 19 | 22 | 20 | 22 |
| Ogden-Clearfield, UT Metro Area | 20 | 21 | 27 | 34 | 28 |
| Boise City, ID Metro Area | 21 | 22 | 34 | 50 | 53 |
| San Luis Obispo-Paso Robles-Arroyo Grande, CA Metro Area | 22 | 14 | 17 | 23 | 24 |
| Des Moines-West Des Moines, IA Metro Area | 23 | 20 | 24 | 27 | 36 |
| Gainesville, FL Metro Area | 24 | 18 | 19 | 17 | 13 |
| Decatur, IL Metro Area | 25 | 35 | 43 | 41 | 42 |
| Oxnard-Thousand Oaks-Ventura, CA Metro Area | 26 | 28 | 26 | 30 | 35 |
| Bismarck, ND Metro Area | 27 | 109 | 153 | 139 | 140 |
| Pittsfield, MA Metro Area | 28 | 23 | 16 | 24 | 21 |
| Rochester, NY Metro Area | 29 | 29 | 33 | 32 | 37 |
| Austin-Round Rock, TX Metro Area | 30 | 33 | 28 | 31 | 32 |
| Athens-Clarke County, GA Metro Area | 31 | 24 | 40 | 40 | 33 |
| Hartford-West Hartford-East Hartford, CT Metro Area | 32 | 26 | 32 | 22 | 26 |
| Salt Lake City, UT Metro Area | 33 | 31 | 30 | 25 | 27 |
| Reno, NV Metro Area | 34 | 36 | 37 | 45 | 51 |
| Peoria, IL Metro Area | 35 | 27 | 20 | 16 | 18 |
| Rapid City, SD Metro Area | 36 | 77 | 120 | 127 | 120 |
| Roanoke, VA Metro Area | 37 | 25 | 38 | 19 | 23 |
| Seattle-Tacoma-Bellevue, WA Metro Area | 38 | 38 | 41 | 35 | 38 |
| Naples-Immokalee-Marco Island, FL Metro Area | 39 | 30 | 29 | 36 | 34 |
| Erie, PA Metro Area | 40 | 42 | 36 | 38 | 41 |
| Springfield, IL Metro Area | 41 | 48 | 68 | 42 | 50 |
| Syracuse, NY Metro Area | 42 | 39 | 44 | 56 | 56 |
| Green Bay, WI Metro Area | 43 | 57 | 63 | 70 | 63 |
| Greeley, CO Metro Area | 44 | 44 | 42 | 37 | 25 |
| Washington-Arlington-Alexandria, DC-VA-MD-WV Metro Area | 45 | 50 | 47 | 47 | 45 |
| San Francisco-Oakland-Hayward, CA Metro Area | 46 | 51 | 55 | 49 | 49 |
| Flagstaff, AZ Metro Area | 47 | 34 | 58 | 54 | 58 |
| Portland-South Portland, ME Metro Area | 48 | 45 | 51 | 51 | 46 |
| South Bend-Mishawaka, IN-MI Metro Area | 49 | 37 | 31 | 26 | 30 |
| York-Hanover, PA Metro Area | 50 | 54 | 69 | 60 | 62 |
| Davenport-Moline-Rock Island, IA-IL Metro Area | 51 | 65 | 61 | 79 | 75 |
| Winchester, VA-WV Metro Area | 52 | 49 | 48 | 44 | 44 |
| Canton-Massillon, OH Metro Area | 53 | 60 | 71 | 77 | 85 |
| Lansing-East Lansing, MI Metro Area | 54 | 66 | 57 | 63 | 60 |
| Lexington-Fayette, KY Metro Area | 55 | 46 | 50 | 43 | 39 |
| Worcester, MA-CT Metro Area | 56 | 53 | 62 | 57 | 57 |
| Santa Rosa, CA Metro Area | 57 | 52 | 60 | 46 | 43 |
| Akron, OH Metro Area | 58 | 59 | 53 | 59 | 52 |
| Harrisonburg, VA Metro Area | 59 | 58 | 127 | 55 | 84 |
| Dover, DE Metro Area | 60 | 124 | 94 | 89 | 93 |
| Colorado Springs, CO Metro Area | 61 | 64 | 73 | 66 | 69 |
| St. Joseph, MO-KS Metro Area | 62 | 41 | 35 | 48 | 138 |
| Sioux City, IA-NE-SD Metro Area | 63 | 40 | 65 | 83 | 73 |
| San Diego-Carlsbad, CA Metro Area | 64 | 61 | 59 | 52 | 48 |
| Milwaukee-Waukesha-West Allis, WI Metro Area | 65 | 75 | 70 | 74 | 74 |
| Rockford, IL Metro Area | 66 | 69 | 67 | 64 | 78 |
| Mount Vernon-Anacortes, WA Metro Area | 67 | 56 | 49 | 61 | 29 |
| Salinas, CA Metro Area | 68 | 79 | 54 | 58 | 54 |
| Fort Wayne, IN Metro Area | 69 | 71 | 75 | 68 | 70 |
| Lancaster, PA Metro Area | 70 | 68 | 66 | 67 | 61 |
| Utica-Rome, NY Metro Area | 71 | 81 | 78 | 125 | 83 |
| Cincinnati, OH-KY-IN Metro Area | 72 | 76 | 76 | 73 | 67 |
| Spokane-Spokane Valley, WA Metro Area | 73 | 55 | 74 | 76 | 89 |
| Denver-Aurora-Lakewood, CO Metro Area | 74 | 72 | 64 | 62 | 64 |
| Huntsville, AL Metro Area | 75 | 70 | 52 | 53 | 55 |
| Bangor, ME Metro Area | 76 | 83 | 56 | 78 | 68 |
| Olympia-Tumwater, WA Metro Area | 77 | 80 | 45 | 71 | 77 |
| Vallejo-Fairfield, CA Metro Area | 78 | 74 | 86 | 84 | 76 |
| Richmond, VA Metro Area | 79 | 84 | 83 | 80 | 81 |
| Palm Bay-Melbourne-Titusville, FL Metro Area | 80 | 78 | 72 | 65 | 40 |
| Buffalo-Cheektowaga-Niagara Falls, NY Metro Area | 81 | 85 | 79 | 81 | 86 |
| Savannah, GA Metro Area | 82 | 73 | 97 | 86 | 88 |
| Bowling Green, KY Metro Area | 83 | 88 | 143 | 146 | 144 |
| Toledo, OH Metro Area | 84 | 92 | 89 | 90 | 105 |
| Dayton, OH Metro Area | 85 | 82 | 82 | 75 | 72 |
| Baltimore-Columbia-Towson, MD Metro Area | 86 | 89 | 84 | 82 | 80 |
| Salem, OR Metro Area | 87 | 94 | 95 | 111 | 126 |
| Omaha-Council Bluffs, NE-IA Metro Area | 88 | 87 | 80 | 72 | 82 |
| Billings, MT Metro Area | 89 | 47 | 85 | 123 | 79 |
| Elizabethtown-Fort Knox, KY Metro Area | 90 | 67 | 116 | 98 | 110 |
| Kansas City, MO-KS Metro Area | 91 | 91 | 88 | 99 | 99 |
| Lebanon, PA Metro Area | 92 | 62 | 81 | 91 | 91 |
| North Port-Sarasota-Bradenton, FL Metro Area | 93 | 95 | 92 | 110 | 92 |
| Salisbury, MD-DE Metro Area | 94 | 99 | 166 | 165 | 170 |
| Atlanta-Sandy Springs-Roswell, GA Metro Area | 95 | 103 | 91 | 96 | 97 |
| Sacramento--Roseville--Arden-Arcade, CA Metro Area | 96 | 102 | 101 | 106 | 103 |
| Reading, PA Metro Area | 97 | 98 | 98 | 94 | 98 |
| Sebastian-Vero Beach, FL Metro Area | 98 | 63 | 39 | 33 | 19 |
| St. Louis, MO-IL Metro Area | 99 | 100 | 87 | 87 | 94 |
| Crestview-Fort Walton Beach-Destin, FL Metro Area | 100 | 86 | 150 | 158 | 139 |
| Missoula, MT Metro Area | 101 | 90 | 46 | 39 | 31 |
| Kingsport-Bristol-Bristol, TN-VA Metro Area | 102 | 118 | 104 | 114 | 123 |
| Springfield, MA Metro Area | 103 | 101 | 90 | 97 | 87 |
| Virginia Beach-Norfolk-Newport News, VA-NC Metro Area | 104 | 97 | 93 | 92 | 100 |
| Orlando-Kissimmee-Sanford, FL Metro Area | 105 | 104 | 96 | 101 | 90 |
| Spartanburg, SC Metro Area | 106 | 115 | 134 | 124 | 131 |
| Houston-The Woodlands-Sugar Land, TX Metro Area | 107 | 113 | 102 | 107 | 104 |
| Detroit-Warren-Dearborn, MI Metro Area | 108 | 114 | 106 | 109 | 102 |
| Albuquerque, NM Metro Area | 109 | 108 | 113 | 112 | 113 |
| Longview, WA Metro Area | 110 | 125 | 163 | 150 | 149 |
| Miami-Fort Lauderdale-West Palm Beach, FL Metro Area | 111 | 116 | 108 | 113 | 107 |
| Wilmington, NC Metro Area | 112 | 112 | 100 | 88 | 101 |
| San Antonio-New Braunfels, TX Metro Area | 113 | 119 | 111 | 108 | 116 |
| Nashville-Davidson--Murfreesboro--Franklin, TN Metro Area | 114 | 121 | 110 | 117 | 125 |
| Springfield, OH Metro Area | 115 | 135 | 125 | 103 | 108 |
| Jacksonville, FL Metro Area | 116 | 117 | 122 | 119 | 111 |
| Johnstown, PA Metro Area | 117 | 96 | 77 | 95 | 127 |
| Deltona-Daytona Beach-Ormond Beach, FL Metro Area | 118 | 129 | 149 | 149 | 151 |
| Lima, OH Metro Area | 119 | 160 | 170 | 169 | 145 |
| Tampa-St. Petersburg-Clearwater, FL Metro Area | 120 | 126 | 117 | 116 | 119 |
| Los Angeles-Long Beach-Anaheim, CA Metro Area | 121 | 123 | 123 | 121 | 124 |
| Medford, OR Metro Area | 122 | 105 | 115 | 69 | 66 |
| Knoxville, TN Metro Area | 123 | 128 | 99 | 104 | 117 |
| Joplin, MO Metro Area | 124 | 111 | 109 | 115 | 130 |
| Longview, TX Metro Area | 125 | 140 | 128 | 131 | 136 |
| Las Vegas-Henderson-Paradise, NV Metro Area | 126 | 130 | 129 | 122 | 115 |
| Memphis, TN-MS-AR Metro Area | 127 | 132 | 130 | 132 | 132 |
| Waco, TX Metro Area | 128 | 150 | 126 | 145 | 142 |
| Grand Junction, CO Metro Area | 129 | 120 | 105 | 120 | 114 |
| Birmingham-Hoover, AL Metro Area | 130 | 138 | 135 | 138 | 137 |
| Victoria, TX Metro Area | 131 | 172 | 137 | 105 | 128 |
| Tulsa, OK Metro Area | 132 | 134 | 118 | 118 | 118 |
| Great Falls, MT Metro Area | 133 | 110 | 174 | 100 | 65 |
| Las Cruces, NM Metro Area | 134 | 137 | 147 | 157 | 169 |
| Chico, CA Metro Area | 135 | 127 | 136 | 142 | 150 |
| Topeka, KS Metro Area | 136 | 143 | 121 | 137 | 155 |
| Amarillo, TX Metro Area | 137 | 144 | 140 | 140 | 129 |
| Oklahoma City, OK Metro Area | 138 | 141 | 133 | 134 | 135 |
| Cape Coral-Fort Myers, FL Metro Area | 139 | 136 | 124 | 128 | 112 |
| Pensacola-Ferry Pass-Brent, FL Metro Area | 140 | 147 | 141 | 136 | 122 |
| Lakeland-Winter Haven, FL Metro Area | 141 | 131 | 139 | 141 | 133 |
| Panama City, FL Metro Area | 142 | 107 | 112 | 135 | 106 |
| Tallahassee, FL Metro Area | 143 | 133 | 103 | 93 | 95 |
| Beaumont-Port Arthur, TX Metro Area | 144 | 148 | 155 | 154 | 148 |
| Stockton-Lodi, CA Metro Area | 145 | 149 | 138 | 143 | 141 |
| Little Rock-North Little Rock-Conway, AR Metro Area | 146 | 145 | 142 | 144 | 147 |
| Yuma, AZ Metro Area | 147 | 157 | 146 | 162 | 160 |
| Anchorage, AK Metro Area | 148 | 146 | 131 | 133 | 143 |
| Bakersfield, CA Metro Area | 149 | 155 | 156 | 160 | 156 |
| Riverside-San Bernardino-Ontario, CA Metro Area | 150 | 151 | 151 | 156 | 157 |
| Ocala, FL Metro Area | 151 | 142 | 145 | 147 | 165 |
| Springfield, MO Metro Area | 152 | 156 | 152 | 164 | 158 |
| Redding, CA Metro Area | 153 | 139 | 158 | 159 | 163 |
| Fort Smith, AR-OK Metro Area | 154 | 163 | 167 | 151 | 154 |
| Corpus Christi, TX Metro Area | 155 | 153 | 148 | 148 | 134 |
| Modesto, CA Metro Area | 156 | 159 | 161 | 153 | 153 |
| Clarksville, TN-KY Metro Area | 157 | 154 | 164 | 161 | 152 |
| Altoona, PA Metro Area | 158 | 106 | 119 | 126 | 96 |
| Jackson, MS Metro Area | 159 | 158 | 165 | 163 | 164 |
| Columbus, GA-AL Metro Area | 160 | 152 | 154 | 152 | 159 |
| Madera, CA Metro Area | 161 | 161 | 144 | 130 | 121 |
| Warner Robins, GA Metro Area | 162 | 175 | 178 | 173 | 162 |
| Fresno, CA Metro Area | 163 | 164 | 159 | 155 | 161 |
| Houma-Thibodaux, LA Metro Area | 164 | 122 | 107 | 85 | 109 |
| Yakima, WA Metro Area | 165 | 162 | 160 | 174 | 166 |
| Lawton, OK Metro Area | 166 | 174 | 162 | 129 | 146 |
| Fairbanks, AK Metro Area | 167 | 178 | 173 | 168 | 168 |
| Yuba City, CA Metro Area | 168 | 170 | 132 | 166 | 171 |
| Albany, GA Metro Area | 169 | 166 | 114 | 102 | 59 |
| Hanford-Corcoran, CA Metro Area | 170 | 167 | 157 | 178 | 178 |
| Visalia-Porterville, CA Metro Area | 171 | 165 | 168 | 167 | 167 |
| Cheyenne, WY Metro Area | 172 | 43 | 18 | 15 | 8 |
| Brownsville-Harlingen, TX Metro Area | 173 | 168 | 172 | 172 | 172 |
| Laredo, TX Metro Area | 174 | 177 | 176 | 176 | 175 |
| Merced, CA Metro Area | 175 | 169 | 171 | 175 | 173 |
| Farmington, NM Metro Area | 176 | 176 | 169 | 171 | 176 |
| El Centro, CA Metro Area | 177 | 171 | 175 | 170 | 174 |
| McAllen-Edinburg-Mission, TX Metro Area | 178 | 173 | 177 | 177 | 177 |

***Table D1:*** **Rank ordering of American MSAs based on their Fitness over the period 2011-2015.**